\documentclass[amsmath,amssymb,
 reprint,%
]{revtex4-1}

\usepackage{graphicx}
\usepackage{dcolumn}
\usepackage{bm}

\usepackage[utf8]{inputenc}
\usepackage[T1]{fontenc}
\usepackage{mathptmx}
\usepackage{etoolbox}

\usepackage{bm}
\usepackage{graphicx}
\usepackage{dsfont}

\usepackage{mathrsfs}

\usepackage{framed}
\usepackage{multirow}
\usepackage[all]{xy}

\usepackage{framed}
\usepackage{comment}
\usepackage{here}                    
\usepackage{latexsym}		     
\usepackage{amsmath}     
\usepackage{amsfonts}  
\usepackage{amssymb}
\usepackage{amsthm}
\usepackage{dcolumn}
\usepackage{color}
\usepackage{mathrsfs}
\usepackage{dsfont}

\newtheorem{definition}{Definition}

\renewcommand{\>}{\rangle}
\newcommand{\<}{\langle}

\newcommand{\ket}[1]{|#1\>}
\newcommand{\bra}[1]{\<#1|}
\newcommand{\braket}[2]{\langle #1 | #2 \rangle}

\newcommand{\be}{\begin{equation}}
\newcommand{\ee}{\end{equation}}
\newcommand{\bea}{\begin{eqnarray}}
\newcommand{\eea}{\end{eqnarray}}

\def\unity{\openone}
\newcommand{\Int}{\mathbb{Z}}

\def\unity{\mbox{\rm\bf 1}}

\renewcommand{\unity}{\mathds{1}}

\usepackage[colorlinks,citecolor=blue,linkcolor=blue,urlcolor=blue]{hyperref}

\makeatletter
\def\@email#1#2{%
 \endgroup
 \patchcmd{\titleblock@produce}
  {\frontmatter@RRAPformat}
  {\frontmatter@RRAPformat{\produce@RRAP{*#1\href{mailto:#2}{#2}}}\frontmatter@RRAPformat}
  {}{}
}%
\makeatother

\begin{document}

\preprint{AIP/123-QED}

\title[Quantum Diffie-Hellman Key Exchange]{Quantum Diffie-Hellman Key Exchange}
\author{Georgios M. Nikolopoulos}
 \email{nikolg@iesl.forth.gr}
\affiliation{ 
Institute of Electronic Structure and Laser, Foundation for Research and Technology-Hellas, GR-70013 Heraklion, Greece
}%

\date{\today}

\begin{abstract}
Diffie-Hellman key exchange plays a crucial in conventional cryptography, as it allows two legitimate users to establish a common, usually ephemeral, secret key. 
Its security relies on the discrete-logarithm problem, which is considered to be a mathematical one-way function, while the final key is formed by random independent actions of the two users. 
In the present work we investigate the extension of Diffie-Hellman key exchange 
to the quantum setting, where the two legitimate users exchange independent random quantum states. 
The proposed protocol relies on the bijective mapping of integers onto a set of symmetric coherent states, and we investigate the regime of parameters for which the map behaves as a quantum one-way function. 
Its security is analyzed in the framework of minimum-error-discrimination and photon-number-splitting attacks, while 
its performance and the challenges in a possible realization are also discussed. 
\end{abstract}

\maketitle

\section{Introduction}
\label{sec1}

In conventional cryptography one typically deals with secret keys that have different lifetimes, depending on their uses and the desired level of security. Typically, the 
lifetime of long-term keys (also known as static or archived) varies from days to  years depending on what they are used for. 
On the contrary, short-term (also known as single-use or ephemeral) keys, are used only for a single sessions or transaction.  Hence, ephemeral keys are mainly 
used for communication, not for storage, and they are deleted at the end of 
the communication session \cite{book1,book2}. 

The  Diffie-Hellman (DH) key exchange plays a pivotal role in the 
generation and the distribution of ephemeral keys \cite{book1,book2}.
By construction,  the protocol does not use any long-term secret key, and its security relies on 
the numerical problem of discrete logarithm, which is a mathematical one-way function. In its simplest form, the protocol involves 
two parties Alice and Bob, who agree publicly on a prime number $p\gg 1$ and its  primitive root, say $g$, modulo $p$. Alice chooses at random an integer 
$a\in[1,p-2]$ and sends to Bob $A = g^a({\rm mod}~p)$. Bob chooses at random an integer $b\in[1,p-2]$ and sends to Alice $B = g^b({\rm mod}~p)$. 
Alice computes $K=B^a ({\rm mod}~p)= (g^{b})^a ({\rm mod}~p) = g^{ab} ({\rm mod}~p) $ 
and Bob computes 
$K=A^b (\mod p)= (g^{a})^b ({\rm mod}~p)= g^{ab} ({\rm mod}~p)$. 

The beauty of the scheme is that  two users  end up with a common random 
key $K$,  without ever needing to send the entirety of the common secret across 
the communication channel (only the integers $p, g, g^a$ and $g^b$ are sent in 
the clear). Its operation relies mainly on the following three facts. (i) The primitive element $g$ ensures that any power $g^x ({\rm mod}~p)$, is an element of the set 
$[1,p-1]$. If $g$ is not primitive, then the security of the protocol is reduced. 
(ii) Any two successive exponentiations commute [i.e., $(g^{a})^b = (g^{b})^a$], thereby leading to a common key for Alice and Bob.
(iii) It is computationally easy to compute $g^x\rm mod~p$ for any given $x\in[1,p-2]$, but it is computationally hard for an adversary to deduce  
$x$ from  $p, g, g^x$, and thus it is also hard to compute the exchanged key $K$.  
The basic  version of the DH key exchange described above is also called "anonymous",  because the participants have no identity that can be verified by either party, and thus 
it is susceptible to a man-in-the middle attack.  
In practice, the DH protocol is generally implemented alongside some means of authentication. For instance, digital certificates and a public-key algorithm, such as RSA, can be used for the verification of the identity of each party. 

There have been attempts in the literature, for the extension of the DH key-exchange protocol to a quantum setting \cite{QDH1,QDH2,QDH3}. 
Although these works lack a rigorous security analysis, their security does not seem to rely on a one-way function, but rather on the same physical principles underlying the security of standard quantum key-distribution (QKD) protocols. 
In this sense they can be viewed as variants of standard QKD protocols, rather than extensions of DH key-exchange to the quantum setting. 

As mentioned above, the discrete logarithm has three important properties, which play fundamental role in the operation as well as in the security of the DH key exchange. 
Up to now it is not clear whether there exists a quantum one-way function (QOWF) with similar properties, thereby allowing for the design of a quantum DH (QDH) key-exchange protocol. 
Especially the commutation of successive operations in quantum physics is not something that comes as easily as in number theory. 
The present work aims at answering these questions, through a comprehensive analysis of a QDH key exchange, 
which relies on mapping classical private keys (integers or binary strings) to symmetric coherent states \cite{ACJ06,Nik08,NikIoa09}. 
After discussing the conditions and the parameters for which this map works as a QOWF, i.e., it is easy to perform, but hard to invert, even for quantum computers,  
we analyze the security of the protocol in the framework of minimum-error-discrimination attack \cite{review1,Ban97}, as well as a photon-number-splitting attack. In the proposed protocol, 
a common quantum key between two users is established from (quantum) information that is contributed by each one of them independently. 

The paper is organized as follows. In Sec. \ref{sec2} we discuss the QOWF under consideration, and in Sec. \ref{sec3} we present the QDH protocol and discuss its robustness against various attacks. 
A summary with concluding remarks is given in Sec. \ref{sec4}.

\section{Quantum One-way Function} 
\label{sec2} 

In analogy to its classical counterpart which relies on the 
discrete-logarithm problem, the design of  an information-theoretically secure 
DH key-exchange protocol in the quantum setting requires 
the existence of a QOWF, which is easy to perform, but hard to invert, even for quantum computers. 
Throughout this work,  we consider the following general setup \cite{ACJ06,Nik08,NikIoa09}. 
All users participating in the  protocol agree on a classical description of a set of pure quantum states 
for a $d-$dimensional quantum system:
\begin{eqnarray}
{\mathbb S}_N \equiv \{\ket{\psi_x}~:~x \in \Int_N\},
\end{eqnarray}
with $\Int_N:=\{0,1, \ldots N-1\}$ and $N\gg 1$. 
The quantum state $\ket{\psi_x}$ is fully identified by the integer $x$. 
 The set is publicly known, and  
for any distinct $x$ and $x^\prime$ in $\Int_N$ we have 
\begin{eqnarray}
|\braket{\psi_{x^\prime}}{\psi_x}| < \zeta,
\end{eqnarray}
for some positive constant $\zeta < 1$. Moreover, we assume that for any given  
$x \in \Int_N$, the preparation of  the state $\ket{\psi_x}$ is easy in the sense 
that it can be performed in (quantum-) polynomial time.

The Holevo information $\chi$ \cite{NCbook} 
sets an upper bound on the information that can 
be extracted from the quantum system, when it is prepared in a state 
chosen at random from the set ${\mathbb S}_N$, and it is given by
\begin{subequations}
\bea
\chi = S\left (\hat{\rho} \right ) - \sum_{x\in \Int_N} 
p_{x} S\left ( \hat{\rho}_{x} \right ),
\label{holevoEq1a}
\eea 
where $S(\cdot)$ is the von Neumann entropy,  
\bea
\label{holevoEq1b}
&&\hat{\rho}_{x} := \ket{\psi_{ x} }\bra{\psi_{x} },\\
&&\hat{\rho} := \sum_{{x} \in \Int_N} p_{ x}  \rho_{x} , 
\label{holevoEq1c}
\eea
\end{subequations}
while $p_{x} $ is the probability for the state $\ket{\psi_{ x} }$ to be chosen, with  
$\sum_{x}  p_{x}  = 1$. For pure states, the second term in 
Eq. (\ref{holevoEq1a}) vanishes giving $\chi = S\left (\hat{\rho} \right ) $. 
For the sake of simplicity, form now on we assume equally probable states i.e., 
$p_{ x}  = 1/N$. The Holevo bound does not make any assumptions about the measurement that is performed for deducing $x$ from $\ket{\psi_x}$. 

Let us consider the following scenario. Alice chooses at random 
${x} \in\Int_{N}$, 
she prepares the quantum system in state $\ket{\psi_{x} }$, which is given to Eve. 
Eve's task is to determine ${x} \in\Int_N$. 
When 
\be
\log_2(N) \gg \chi, 
\label{qowf_cond1}
\ee
the bijective map
\begin{eqnarray}
{x} \mapsto \ket{\psi_{ {x} }}
\label{map:eq}
\end{eqnarray}
is a QOWF in the sense that, for a given
${x} \in \Int_N$, the deterministic preparation of the system in the state
$\ket{\psi_{ x} }$ is possible via the classical description of ${\mathbb S}_N$,
while the inversion of the map (with nonnegligible probability) is
guaranteed impossible by the Holevo bound \cite{NCbook}. 
For $\log_2(N) \gg \chi$ the information that Eve can  extract from the system is far less than the $\log_2(N) $ bits required in order to determine fully Alice's random integer $x $. If $ x $ is initially unknown, 
it remains unknown (with high probability) 
even after Eve has access to $\ket{\psi_{ x} }$.  

\subsection{A criterion for a QOWF map}
\label{sec2a}

The condition (\ref{qowf_cond1}) is somewhat vague, leaving space for debate on the difference that 
$\chi$ and $\log_2(N)$ must have in order for the condition to be satisfied. 
For example, should $\chi$ be five or ten times larger than $\log_2(N)$?
A more concrete condition can be derived by noting that the Holevo information is 
related to the average error probability in deducing the integer $x$, which is the 
main task of Eve. For $\chi \neq 0$ 
there is a non-zero probability for Eve to deduce the wrong value of $x$ given 
the quantum state $\ket{\psi_{ x} }$. The average error probability $p_{\rm err} $
obeys the Fano inequality   \cite{NCbook} 
\bea
H(p_{\rm err} )+p_{\rm err} \log_2(N-1)
 &\geq& \log_2(N)-\chi,   
 \label{fano1:eq}
\eea
where $H(\cdot)$ is the binary  entropy. 
Inequality (\ref{fano1:eq}) suggests that it gets harder to determine ${ x} $ (i.e., the error 
probability increases), for decreasing values of $\chi$. 
The Fano inequality is very general and it does not make any assumptions about the measurement  
applied on the given quantum state.  
It is natural to assume, however, that in order to deduce $x$ from the given 
quantum state $\ket{\psi_{ x} }$, Eve will choose the most general measurement, which is allowed
 by the laws of quantum physics, and it is optimal in the sense 
that it minimizes her error probability, and thus maximizes the probability of success. 
Of course, there is always the possibility for Eve to guess correctly the value of ${x} $ without interacting 
at all with the given state, with the corresponding 
probability of success given by 
$p_{\rm suc}^{\rm (rg)}=1/N$. 
The random guess is the best that Eve can do if the map 
${ x} \mapsto \ket{\psi_{  x }}$ 
was an ideal QOWF. In practice, we expect deviations from the ideal QOWF, 
which can be quantified by the ratio  
\bea
D:= \frac{\left | p_{\rm err}^{\rm (min)}  - p_{\rm err}^{\rm (rg)}\right |}{p_{\rm err}^{\rm (rg)}},   
\label{qowf_cond2}
\eea
where $p_{\rm err}^{\rm (rg)}=1-p_{\rm suc}^{\rm (rg)} = (N-1)/N$, while $p_{\rm err}^{\rm (min)} $ is the minimum error probability.
The smaller  $D$ is, the closer the map is to an ideal QOWF. 
For our purposes, we introduce 
a security parameter $\epsilon\ll 1$, and the map will be  
considered to be a QOWF if $D\leq \epsilon$.  
The optimal minimum-error measurement is known 
for special cases only, and in the following we will consider such a case. If such a measurement is not known for the set of states under consideration, one may still  use condition (\ref{qowf_cond2}) by replacing $p_{\rm err}^{\rm (min)}$ 
with the error rate given by the Fano inequality (\ref{fano1:eq}). 

\subsection{Implementation with coherent states}
\label{sec2b}

The set of states under consideration is 
\begin{subequations}
\label{CoheSet:Eqs}
 \begin{eqnarray}
{\mathbb S}_N \equiv \left \{\ket{\psi_x}:=\ket{\sqrt{\mu}e^{{\rm i} x \delta\varphi}}~:~\delta\varphi:=\frac{2\pi}N,~ x \in \Int_N\right \},
\label{set_coh:eq}
\end{eqnarray} 
for even $N>4$, where $\mu$ is the mean number of photons and the single-mode 
coherent state $\ket{\psi_x}$ is eigenstate of the bosonic annihilation operator $\hat{a}$ i.e., 
$\hat{a}\ket{\psi_x} = \psi_x \ket{\psi_x}$. It can be expanded in terms of the Fock states 
$\{\ket{n}\}$ as follows 
\bea
\ket{\psi_x}:=e^{-\mu/2} \sum_{n=0}^\infty \frac{\psi_x^n}{\sqrt{n!}}\ket{n}.
\label{alpha_fock:exp}
\eea
In phase space, all the coherent states in ${\mathbb S}_N$ have the same mean number of photons $\mu$, but their phases are distributed around the circle at regular intervals of $\delta\varphi$.
The unitary transformation which maps each state onto its successor 
is the elementary phase-shift operator 
\bea
\hat{\cal U} = e^{{\rm i} 2\pi \hat{a}^\dag \hat{a}/N} =e^{{\rm i} \delta\varphi \hat{a}^\dag \hat{a}} ,
\eea
where $\hat{a}^\dag$ is the bosonic creation operator. 
Indeed, one can readily show that 
$\hat {\cal U}^\dag \hat{a} \hat {\cal U} = \hat{a} e^{{\rm i}2\pi/N}=\hat{a} e^{{\rm i} \delta\varphi} $, which implies that 
\bea
\hat {\cal U}\ket{\psi_{x \ominus 1}} = 
\ket{\psi_{x\ominus 1} e^{{\rm i}\delta\varphi}} = \ket{\psi_{x}} = \hat{\cal U}^{x}\ket{\psi_{0}} , 
\eea 
where $\hat{\cal U}^{N} = \hat{\unity}$, while $\ominus$ denotes subtraction modulo $N$. 
\end{subequations}

There are many advantages in considering the set of symmetric coherent states 
(\ref{set_coh:eq}) for the design of cryptographic protocols. From a practical point of view, coherent states is a standard information carrier in various quantum communication tasks, including QKD protocols, and there is all the necessary  knowhow for the preparation, the manipulation and the measurement of such states. 
In particular, one can shift the phase of a coherent state using standard phase modulators, thereby encoding easily the desired $x$  \cite{LucYuaDynShi18}. From a theoretical point of view, the symmetry of the states simplifies considerably the calculations. Most importantly, it is well known that the square-root measurement is optimal, in the sense that it minimizes the error probability in deducing $x$, from  an unknown state $\ket{\psi_x}$ 
chosen at random from the set (\ref{set_coh:eq}). 
The associated minimum error probability is given by \cite{review1,Ban97}
\begin{subequations} 
\bea
p_{\rm err}^{\rm (min)}:= 
\sum_{x\in\Int_N} \sum_{y\neq x} p_x P(y|x)
=\frac{1}N \sum_x \sum_{y\neq x}  {\rm Tr}(\hat{\Pi}_y \hat{\rho}_x)
\nonumber\\
\eea
where $y\in\Int_{N}$, 
\bea
\hat{\Pi}_y = \frac{1}N\hat{\rho}^{-1/2}~\hat{\rho}_y~ \hat{\rho}^{-1/2}, 
\eea
\end{subequations} 
and the density operators are given by Eqs. (\ref{holevoEq1b}) and 
(\ref{holevoEq1c}). The conditional 
probability $P(y|x)$ is the probability to obtain outcome 
$y$ given that the measured state is $\ket{\psi_x}$. 
Using the expansion (\ref{alpha_fock:exp}), one can show that 
\bea
\hat{\rho} &=& e^{-\mu} \sum_{n=0}^{\infty}\sum_{n^\prime=0}^{\infty} 
\frac{\mu^{(n+n^\prime)/2}}{\sqrt{n!n^\prime!}}\ket{n}\bra{n^\prime} 
\delta(|n-n^\prime|=q N)
\nonumber\\ 
&=&  \sum_{n=0}^{\infty}\sum_{n^\prime=0}^{\infty} 
\sqrt{{\mathscr P}(\mu,n){\mathscr P}(\mu,n^\prime)}
\ket{n}\bra{n^\prime} 
\delta(|n-n^\prime|=q N), 
\nonumber\\
\eea
for $q\in \mathbb{N}_0$, where $\delta(\cdot)$ has non-zero contributions 
for $|n-n^\prime|=0,N,2N, \ldots$ only, and ${\mathscr P}(\mu,n)$ denotes 
the Poisson distribution. 
The Poisson distribution is discrete and with non-negligible 
probabilities mainly for values of $n$ in the interval $[n_{\rm min}, n_{\rm max}]$, with $n_{\rm max} := \lceil \mu+6\sqrt{\mu} \rceil$ and 
$n_{\rm min} := \max\{ 0,\lfloor \mu-6\sqrt{\mu} \rfloor \}$. 
If $N\geq n_{\rm max} - n_{\rm min}$, then  
the function $\delta(\cdot)$ essentially reduces to Kronecker delta $\delta_{n,n^\prime}$, and the density operator becomes diagonal for all practical purposes  
\bea
\hat{\rho} \simeq e^{-\mu}  \sum_{n=0}^{\infty} 
\frac{\mu^n}{n!}\ket{n}\bra{n} :=
\sum_{n=0}^{\infty} {\mathscr P}(\mu,n)\ket{n}\bra{n}. 
\label{rho_diag:eq}
\eea  
It is worth noting here that when the parameters in the protocol are such that $\hat{\rho}$  reduces to Eq. (\ref{rho_diag:eq}), then it also becomes independent of the phase slices; a property which is also expected to be reflected at various measures to be discussed in the following.

\begin{figure}
\centerline{\includegraphics[width=1\linewidth]{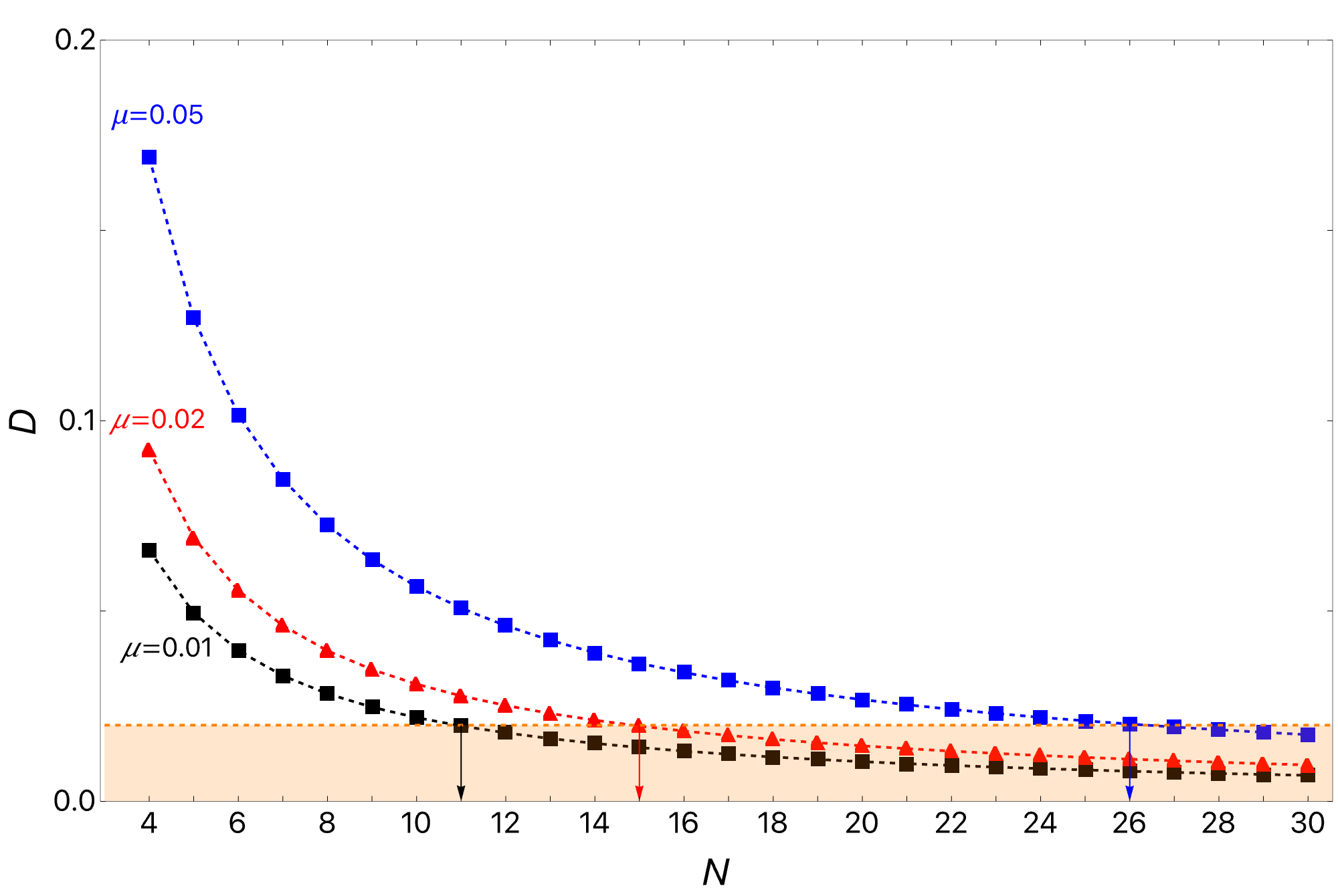}}
\caption{(Color online)  The relative difference (\ref{qowf_cond2}), as a function of the number of symmetric coherent states $N$, for three different values of the mean number of photons $\mu$. The minimum-error probability is obtained for the 
square-root measurement, as described in the text. 
The dashed horizontal line marks the region (orange area) where $D\leq \epsilon$ for $\epsilon = 2\times 10^{-2}$. The vertical arrows point at the 
value of $N$, above which the curve for a given $\mu$ enters the area of  
$D\leq \epsilon$, and the map $x\mapsto\ket{\psi_x}$ 
is considered to be close to an ideal QOWF.}
\label{D:fig}
\end{figure}

At this stage we have all the necessary ingredients in order to investigate 
the regime of parameters for which the map $x\mapsto \ket{\psi_x}$ may serve as a QOWF.  
In Fig. \ref{D:fig} we plot the ratio (\ref{qowf_cond2}) 
as a function of the number of phase slices $N$, 
for different values of the mean number of photons $\mu$. 
For a given $\mu$, the difference decreases monotonically  with increasing values of $N$, and it crosses the chosen $\epsilon = 2\times 10^{-2}$ at some value, say $N=N^\star$, which depends on the mean number of photons.  
More precisely, $N^\star$ increases with increasing 
$\mu$, which implies that one needs more states in the 
set $\mathbb S_N$, in order to ensure that the map $x\mapsto \ket{\psi_x}$ operates as a QOWF, based solely on the criterion $D\leq\epsilon$. 
The fact that, for a given $\mu$, the map gets closer 
to the ideal QOWF as we increase $N$, ensures that it is getting harder for an adversary to deduce $x$ from a given randomly chosen state from the 
set (\ref{set_coh:eq}). Indeed, the overlap of successive states increases with decreasing $\delta\varphi$ as
\bea
|\< \psi_x\ket{\psi_{x\pm 1}}| 
&\simeq&e^{-\mu \delta\varphi^2/2},~\textrm{for}~N\geq 4. 
\eea

\begin{figure}
\centerline{\includegraphics[width=1\linewidth]{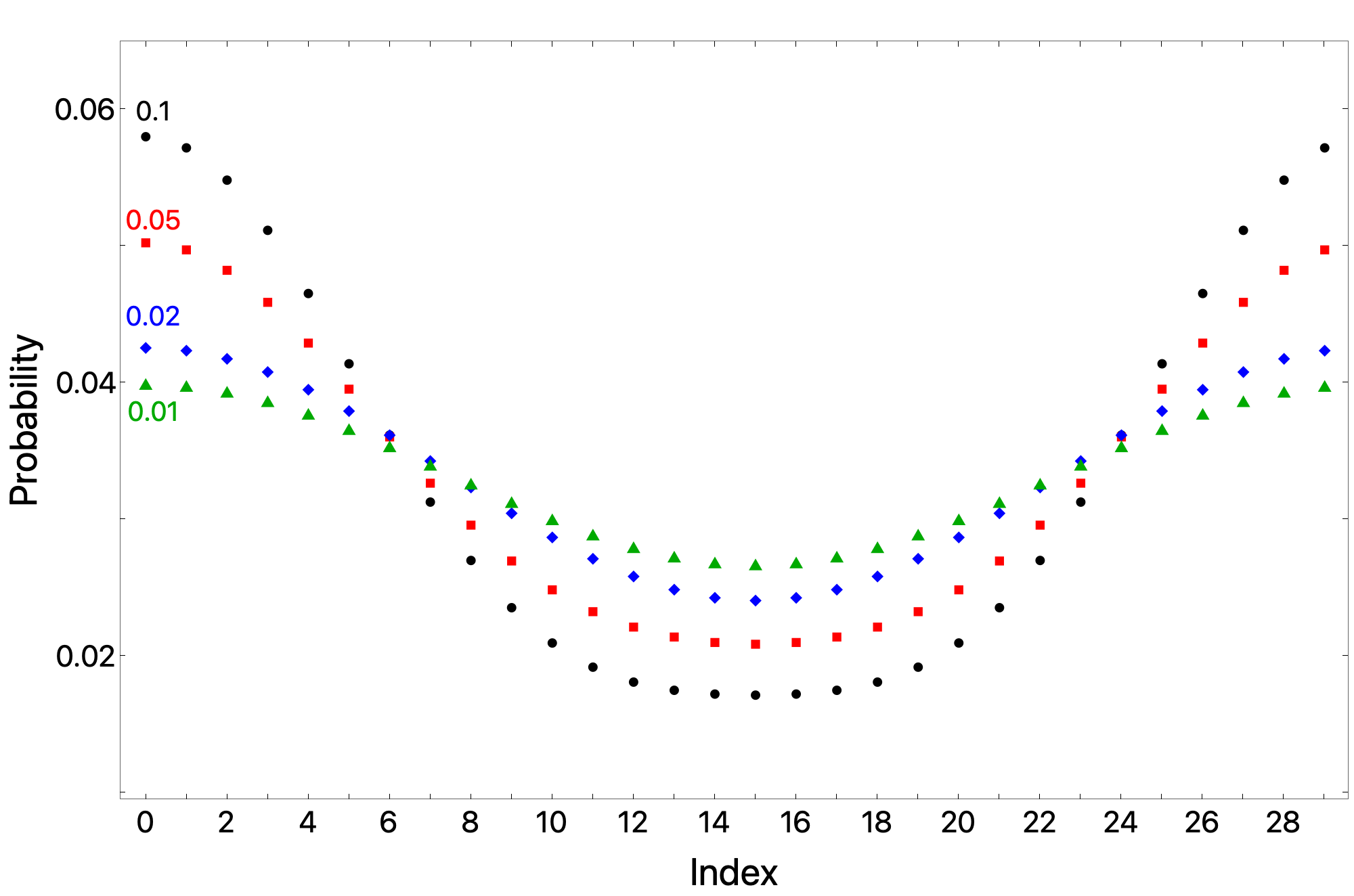}}
\caption{(Color online)  Conditional probability $P(y|x)$  as a function of $y\in\Int_{30}$, for $x=0$, and various values of $\mu$. }
\label{Pcond:fig}
\end{figure}

The probability for Eve to obtain outcome $y$ given that the 
input state is $\ket{\psi_{x}}$ is given by 
$P(y|x) = {\rm Tr}(\hat{\Pi}_y \hat{\rho}_{x})$. 
As depicted in Fig. \ref{Pcond:fig}, 
this probability exhibits a maximum at $y=x$, and 
it gets smaller for indices $y\neq x$. 
The larger $\mu$ is the more  prominent the peak of the distribution becomes, which reflects that Eve's capability of deducing the right state by means of the minimum-error-discrimination measurement improves as we increase the mean number of photons, for given $N$.

One may also look at the information gain from Eve's point of view 
\bea
{\rm Gain} := H(x) - H(x|y), 
\eea 
where $x$ and $y$ are random variables representing Alice's choice and Eve's outcome, respectively. For equally probable states in the set (\ref{set_coh:eq}), we have $H(x) = \log_2(N)$, whereas 
$H(x|y)$ is the post-measurement entropy and can be readily calculated from $P(y|x)$ and 
Bayess law. 
Our simulations suggest that, due to symmetry [recall Eq. (\ref{rho_diag:eq})], the gain does not vary appreciably with $N$ for $N>10$, and some values are shown in table \ref{tab1}, for  different combinations of $\mu$, and $N$. 
Note that for the particular choice of parameters, $D\leq \epsilon$ for all values of $\mu$ shown in the table, 
and in this sense the map (\ref{map:eq}) operates as a QOWF (see Fig. \ref{D:fig}). 
For a given $\mu$, the entropy $H(x)$ increases with increasing $N$, but $H(x|y)$ also increases accordingly, so that the gain remains the same (not shown here). Moreover, as expected, 
for given $N$ the gain increases with increasing $\mu$.
The information gain is found to be considerably (at least 30 times) smaller than $H(x)$, or equivalently $H(x)\simeq H(x|y)$, 
which suggests that, in the particular regime of parameters, Eve's ignorance after the minimum-error discrimination measurement, remains practically the same as before any measurement. 
This is also another manifestation of the closeness of the map to 
the ideal QOWF, for the parameters shown in Fig. \ref{D:fig} and table \ref{tab1}. 

\begin{table}
\centering
\caption{Entropies and information gain, for the minimum-error-discrimination measurement and the set of states (\ref{set_coh:eq}). }
\label{tab1}
\begin{tabular}{c|c|c|c|c|c}
\hline
\hline
$\mu$ ~& ~$N$~ &~ $H(x)$~  &~ $H(x|y)$ ~&~Gain ~ &~$\chi$\\ \hline
0.01    ~& ~ 20 ~  & ~ 4.322 ~ &~  4.307  ~  &~ 0.014 ~ & ~0.080 ~\\ 
  0.02    ~& ~ 20 ~ & ~ 4.322  ~ & ~ 4.293  ~ &~  0.029 ~ & ~0.139 ~\\ 
  0.05    ~& ~ 30 ~ & ~ 4.907 ~ & ~ 4.836  ~  &~  0.071 ~ & ~0.289 ~\\ 
  0.1    ~& ~ 40 ~  & ~ 5.322 ~ &~  5.182  ~  &~ 0.140 ~ & ~0.480 ~\\ 
 \hline
 \hline
\end{tabular}
\end{table}


\section{A Quantum Diffie-Hellman Protocol}
\label{sec3}

The QOWF discussed in the previous section is the starting 
point for the generalization of the DH protocol to the quantum setting. 
As discussed in Sec. \ref{sec1}, besides the one-way character of the map under consideration, there are two more key-points 
in the classical DH protocol, i.e., the use of a primitive element in the set of integers, and  the commutation of successive exponentiations.    
One way to transfer these two important features to the quantum setting 
is to consider a set of equally probable symmetric coherent states governed by 
Eqs. (\ref{CoheSet:Eqs}). We also have that 
\bea
\left [ \hat{\cal U}^x, \hat{\cal U}^y \right ] = 0, 
\label{commut}
\eea
  and 
 \bea
 \hat{\cal U}^x \hat{\cal U}^y =  \hat{\cal U}^{x\oplus y},  
 \label{prodU}
 \eea
 for any $x,y\in\Int_N$, where $\oplus$ denotes addition modulo $N$. 

\begin{figure}
\centerline{\includegraphics[width=0.9\linewidth]{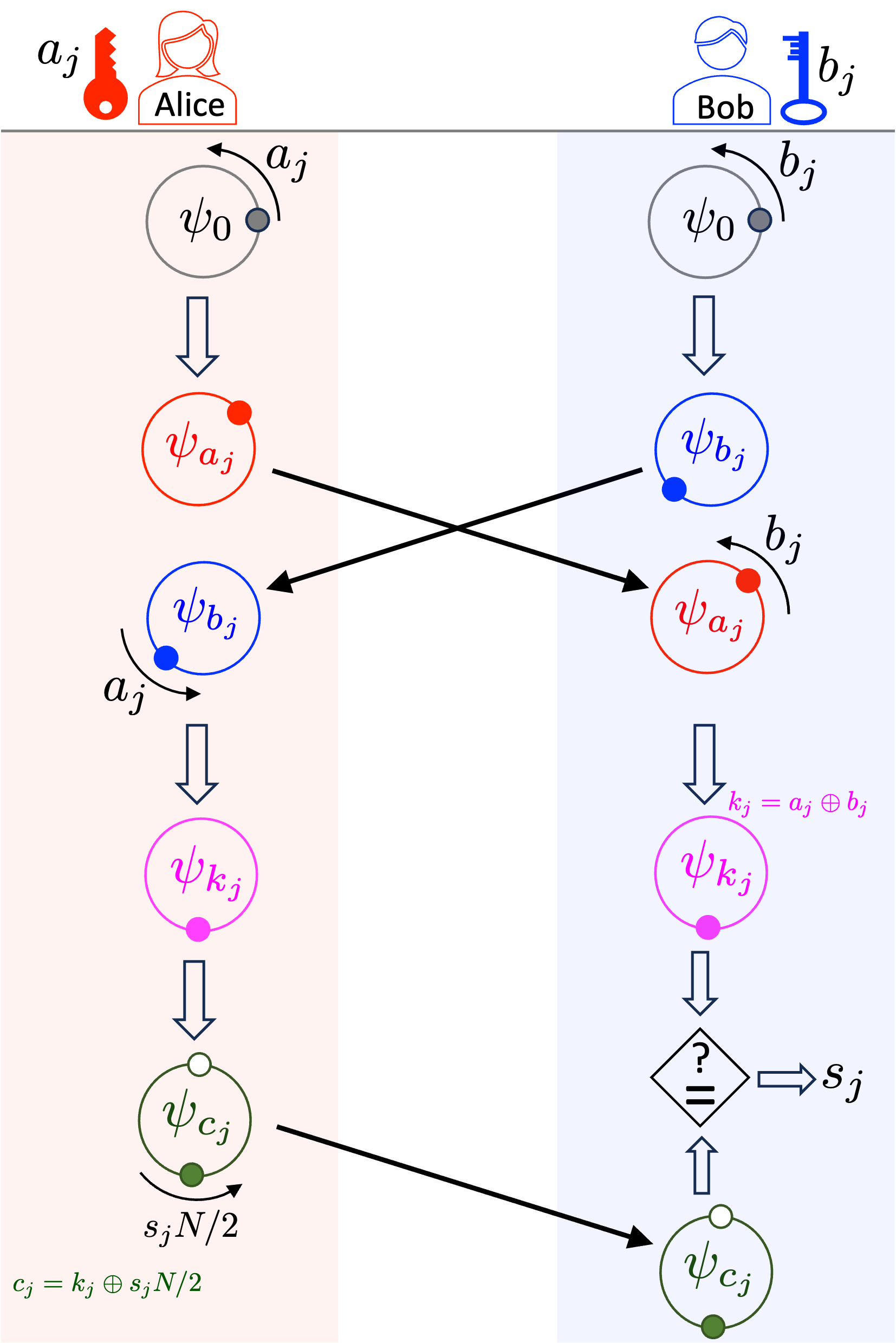}}
\caption{(Color online)  Schematic representation of the quantum Diffie-Hellman key-exchange scheme under consideration.}
\label{QDH:fig}
\end{figure}

Consider now two parties Alice and Bob, who are far apart (at a fixed distance)
and they want to establish a common secret key. 
They are connected by a reliable quantum channel (i.e., losses and phase drift are sufficiently low so that the quantum states that are exchanged between 
Alice and Bob will arrive their destination with a non-negligible probability, and without being totally randomized),  
a classical authenticated service channel, while they have a common phase reference. 
The set of states ${\mathbb S}_N$ as well as the underlying unitary operator ${\hat{\cal U}}$ are publicly known. 
A quantum version of the DH key-exchange protocol is outlined in Fig. \ref{QDH:fig},  
and the main steps are the following. 
\begin{definition} Quantum Diffie-Hellman key agreement (basic version).
\begin{enumerate}
\item {\rm Alice chooses at random a private key $a_j\in\Int_N$ and prepares 
the quantum state $\ket{\psi_{a_j}} = \hat{\cal U}^{a_j} \ket{\psi_0}$ which is sent to Bob. }
\item {\rm Bob chooses at random a private key $b_j\in\Int_N$ and prepares 
the quantum state $\ket{\psi_{b_j}} = \hat{\cal U}^{b_j} \ket{\psi_{0}}$ which is sent to Alice. }
\item {\rm Bob applies $\hat{\cal U}^{b_j}$ on the received state thereby obtaining the quantum key $\ket{\psi_{k_j}} = \hat{\cal U}^{b_j}\ket{\psi_{a_j}}$, with $k_j=b_j\oplus a_j$.}
\item   {\rm Alice applies $\hat{\cal U}^{a_j}$ on the received state thereby obtaining the quantum key $\ket{\psi_{k_j}} =\hat{\cal U}^{a_j}\ket{\psi_{b_j}}$, with $k_j=a_j\oplus b_j$.}
\item Alice chooses a random bit $s_j\in\{0,1\}$ and encodes it on the quantum key $\ket{\psi_{k_j}} $ as follows
\bea
\hat{\cal U}^{s_j N/2}\ket{\psi_{k_j}}  := \ket{\psi_{c_j}} ,
\label{encoding:eq}
\eea 
where $c_j = k_j\oplus s_jN/2$. The cipher-state $\ket{\psi_{{c}_j}}$  is sent to Bob. 
\item Bob interferes the cipher-state  $\ket{\psi_{c_j}}$ with his quantum key   
$\ket{\psi_{k_j}}$ on a 50:50 beam splitter with the two output ports monitored by two detectors $D_{0}$ and $D_{1}$, in order to deduce $s_j$. 
When only detector $D_{l}$ clicks, he concludes $s_j = l$, and records an inconclusive outcome otherwise.
\item The steps 1-6 are repeated many $(M\gg 1)$ times and all of the inconclusive outcomes are discarded. 
\item Alice and Bob choose randomly a sufficiently large number of bits, to be sacrificed in order to decide whether eavesdropping took place 
during the transmission. If eavesdropping is likely to have taken place, they abort the protocol otherwise they proceed to the next step. 
\item Alice and Bob extract  a common binary secret key from the remaining bits, by applying error correction and privacy amplification, as discussed in the following subsections. 
\end{enumerate}
\end{definition} 

By virtue of Eq. (\ref{commut}),  in an ideal scenario at the end of the protocol Alice and Bob share the same quantum key, because 
$\ket{\psi_{k_j}} = \hat{\cal U}^{b_j}\ket{\psi_{a_j}} = 
\hat{\cal U}^{b_j}\hat{\cal U}^{a_j} \ket{\psi_{0}} = 
\hat{\cal U}^{a_j}\hat{\cal U}^{b_j} \ket{\psi_{0}} = 
\hat{\cal U}^{a_j}\ket{\psi_{b_j}}$. 
Note that this holds because Alice and Bob begin with the same initial state $\ket{\psi_{0}}$, which is possible if they share a common phase reference. 
A nice property of the protocol is that the final quantum 
key has been formed by the random and independent actions of both users. 
This means that neither Alice or Bob know the shared quantum state, because the 
random phase-shift applied by the other user is random and uniformly distributed 
over $[0,2\pi)$. 
 
\subsection{Extraction of classical key}
\label{sec3a}

As discussed above, at the end of step 4, Alice and Bob share a common secret quantum key $\ket{\psi_{k_j}} $, and neither of them knows the state. 
One may consider various ways for extracting a numerical key from the quantum key. 
For instance, one way for Alice and Bob is to apply independently unambiguous state discrimination or dual-homodyne detection on 
the components of their quantum keys, in order to deduce $k_j$.  
However, these approaches turn out to be very inefficient because, as discussed in the previous section, successive coherent states in the set (\ref{set_coh:eq}) must have 
sufficiently large overlap, in order for the map under consideration to serve as a QOWF. To circumvent this difficulty, in the proposed protocol we follow steps 5 and 6. 

For the sake of simplicity, the protocol above has been outlined for the ideal scenario. To account for possible losses, imperfections, and eavesdropping, 
from now on we introduce additional superscripts $u\in\{{\rm A,B}\}$ for the states of Alice and Bob.
Hence, in step 5 Alice chooses a random bit and encodes it onto her quantum key $\ket{\psi_{k_j^{\rm (A)}}}$,  using the operation  
\bea
\hat{\cal U}^{s_j N/2}\ket{\psi_{k_j^{\rm (A)}}}  := \ket{\psi_{c_j}} ,
\label{encoding:eq}
\eea 
with $c_j = k_j^{\rm (A)}\oplus s_jN/2$.
In order to deduce $s_j$,  Bob interferes 
the $j$th cipher state  $\ket{\psi_{c_j}}$ with his quantum key   
$\ket{\psi_{{k_j}^{\rm (B)}}}$ on a 50:50 beam splitter, 
and the two output modes are monitored by means of two single-photon detectors that we name ${ D}_0$ and ${ D}_1$.  

The states at the two outputs of the beam splitter are  $\ket{\omega_0} \otimes \ket{\omega_1}$ where 
\bea
\omega_{l} &=&\frac{\eta \psi_{c_j}+(-1)^l \sqrt{\eta} \psi_{k_j^{\rm (B)}}}{\sqrt{2}} 
 \nonumber\\
& =&\sqrt{\eta}\psi_{k_j^{\rm (B)}}
\frac{\sqrt{\eta}e^{{\rm i}s_j\pi} e^{{\rm i}\phi} 
+(-1)^l}{\sqrt{2}}, 
\label{omegaL:eq1}
\eea
is the state at output $l\in\{0,1\}$.
The parameter $0\leq \eta\leq 1$ denotes the channel transmissivity, and we have taken into account that the cipher state has experienced these losses twice. 
For typical single-mode fibers used in QKD experiments, $\eta = 10^{-\alpha L/10}$, where $\alpha \simeq 0.2 {\rm dB/km}$ is the linear attenuation and $L$ the length of the fiber.   Moreover, we have introduced 
the random variable $\phi$, which accounts for possible random phase shifts of $\psi_{c_j}$ relative to $\psi_{k_j^{\rm (B)}}$. In order to take into account the finite efficiency of photon detectors, $0<\eta_d<1$, 
we can replace $\sqrt{\eta} \psi_{k_j^{\rm (B)}}$ by $\sqrt{\eta \eta_{\rm d}} \psi_{k_j^{\rm (B)}}$ in the last equation. The probability 
of dark counts in standard commercially available superconducting nanowire single-photon 
detectors is $\sim 10^{-7}$, and it can be ignored in our analysis because 
it is negligible compared to the probability of signal count.

The conditional probability of no click at detector $D_l$  is given by
\bea
&&P(\textrm{no click at }D_l|\phi,s_j) = |\langle 0 | \omega_l\rangle|^2 
\nonumber\\
&=&
\exp \left \{ 
-\frac{\eta\eta_d\mu}2
\left [ 1+\eta-2(-1)^{l+s_j} \sqrt{\eta}\cos(\phi ) \right ] 
\right \}.
\label{PnoDl:eq}
\eea 
Ideally $(\eta=1, \eta_d = 1,  \phi=0)$, we have 
\bea
P_{\rm id}(\textrm{no click at }D_l|s_j) = 
\exp \left \{ 
-\mu\left [ 
 1+(-1)^{s_j+l} \right ] 
\right \},
\eea
which dictates the following decision-making strategy for Bob: 
When only detector $D_l$ clicks, he concludes that $s_j = l$. 

\begin{figure}
\centerline{\includegraphics[width=1\linewidth]{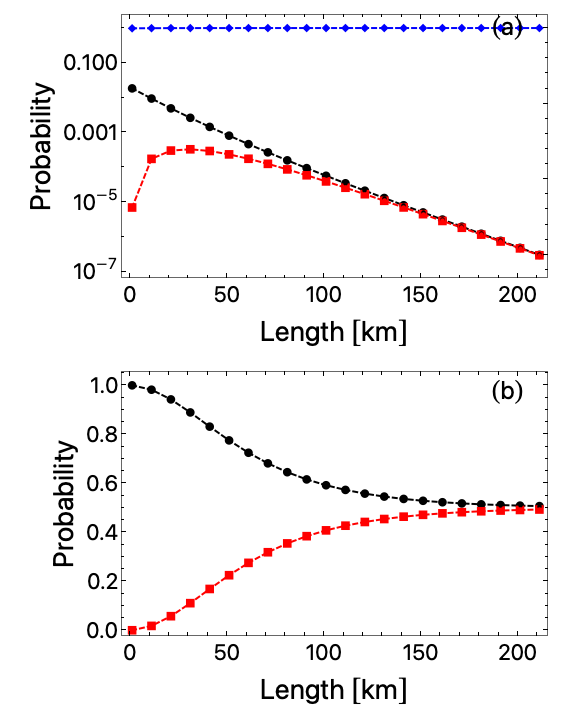}}
\caption{(Color online) The marginal probabilities $P(\textrm{cor})$ (black disks), $P(\textrm{inc})$ (blue diamonds) 
and  $P(\textrm{err})$ (red squares), are plotted as functions of the propagation length $L$, for $\mu = 0.02$, before  (a), and after (b) the rejection of the data pertaining to inconclusive outcomes. Other parameters: $\alpha = 0.2 {\rm dB/km}$, $\eta_{\rm d} = 0.5$, ${\cal D} = 10^{-3} {\rm rad}^2{\rm km}^{-1}$.} 
\label{Pcei:fig}
\end{figure}

Monitoring the outcomes of two detectors Bob can have the following events 
\begin{enumerate}
\item {\em Only  detector $D_l$ clicks}. This event 
occurs with probability 
\bea
P(\textrm{click at }D_l|\phi,s_j) &=& [1-P(\textrm{no click at }D_l|\phi,s_j) ]
\nonumber\\
&\times&P(\textrm{no click at }D_{l\oplus 1}|\phi,s_j),
\eea
and Bob concludes that $s_j = l$. 
Ideally we have
\bea
P_{\rm id} (\textrm{click at }D_l|s_j)= 
e^{-2\mu}\left [-1+e^{\mu[1+(-1)^{l+s_j}]}\right ] ,
\eea
which shows that for $s_j = 0$, only the detector $D_0$ clicks, 
while for $s_j = 1$, only the detector $D_1$ clicks. 
\item {\em None of the detectors clicks}. This is an inconclusive outcome, 
which occurs with probability 
\bea
P(\textrm{no click}|\phi,s_j)  &=& P(\textrm{no click at }D_0|\phi,s_j)
\nonumber\\
&\times&
P(\textrm{no click at }D_1|\phi,s_j)
\eea
Ideally we have $P_{\rm id} (\textrm{no click}|s_j) = e^{-2\mu}$. 
\item {\em Both of the detectors click}. This is also 
an inconclusive outcome, which occurs with the probability 
\bea
P(\textrm{double click}|\phi,s_j) &=& 
 [1-P(\textrm{no click at }D_0|\phi,s_j)]
\nonumber\\
&\times&
[1-P(\textrm{no click at }D_1|\phi,s_j)],
\nonumber \\
\eea
and it pertains to the presence of multiphoton components in the 
coherent states. When $\mu\ll 1$, the presence of multiple 
photons is negligible, and thus the probability of coincidence is expected 
to be significantly smaller than the previous probabilities. 
\end{enumerate}

The overall probability for an inconclusive outcome is 
$P(\textrm{inc}|\phi,s_j) = P(\textrm{no click}|\phi,s_j)+P(\textrm{double click}|\phi,s_j)\simeq P(\textrm{no click}|\phi,s_j)$, which gets smaller 
with increasing $\mu$. On the other hand, having fixed the desired value of $\mu$, 
according to the findings of Sec. \ref{sec2b},  one can choose the number 
phase slices (i.e., the number of states in the set), so that the map behaves as QOWF, thereby ensuring the security of the protocol against a potential adversary.

The random phase drift $\phi$ is expected to have a normal distribution, 
centered at zero, and with standard deviation $\sigma_\phi$ 
i.e., $p(\phi) = {\cal N}(0, \sigma_\phi)$ \cite{MinRieSimZbiGisPRA08,LucYuaDynShi18}.  
The probabilities for the aforementioned events vary with $\phi$. Given that a very large number of random independent states is 
exchanged between Alice and Bob and they experience independent random phase drift, 
we are interested in the marginal probabilities 
\begin{subequations}
\bea
\label{Pcon:eq}
&&P(\textrm{cor})=\frac{1}{2}\sum_{s_j}
\int_{-\infty}^\infty P(\textrm{click at } D_{l=s_j}|\phi,s_j)p(\phi)d\phi
\\
\label{Pinc:eq}
&&P(\textrm{inc})= \frac{1}{2}\sum_{s_j}
\int_{-\infty}^\infty P(\textrm{inc}|\phi,s_j)p(\phi)d\phi
\\
\label{Perr:eq}
&&P(\textrm{err})=\frac{1}{2}\sum_{s_j}
\int_{-\infty}^\infty P(\textrm{click at }D_{l=s_j\oplus 1}|\phi, s_j)p(\phi)d\phi
\nonumber\\
\eea
\end{subequations}
where $P(\textrm{cor})$ is the probability for conclusive correct outcome,  $P(\textrm{inc})$ is the probability for inconclusive outcome, and $P(\textrm{err})$ is the error probability, while we have assumed that the random phase shift is independent of the value of $s_j$. 

Interestingly enough, the probabilities $P(\textrm{cor})$, $P(\textrm{inc})$ and 
$P(\textrm{err})$ do not depend on the number of phase slices $N$. This is  because 
$ \langle 0 | \omega_l\rangle  = e^{-|\omega_l|^2/2}$,
and thus Eq. (\ref{PnoDl:eq}) is independent of $N$. They do depend, however, on the mean 
number of photons $\mu$ and the standard deviation of the phase drift $\sigma_\phi$, which increases with the length of propagation $L$. 
Assuming a phase-diffusion model we have  
$\sigma_\phi\simeq \sqrt{{\cal D}L}$, where ${\cal D}$ is the diffusion coefficient \cite{MinRieSimZbiGisPRA08}. Typically 
${\cal D}\lesssim 10^{-3}{\rm rad}^2{\rm km}^{-1}$ for optical fibers \cite{MinRieSimZbiGisPRA08}, which is in agreement with the various phase-drift rates that have been reported in the literature for different implementations  \cite{LucYuaDynShi18,twin2,twin3,twin4} .
As depicted in Fig. \ref{Pcei:fig}(a), the probability for inconclusive outcome 
is quite large, due to the low mean number of photons, but it does increase with the distance (not visible in the figure).   
At the same time the probability of correct outcome decreases with $L$, 
whereas the error probability does not follow a monotonic behavior. 
Both of them, however, decrease as 
\bea
P_\infty (\textrm{cor})= P_\infty (\textrm{err}) = e^{-\eta ( 1+\eta ) \eta_{\rm d}\mu/2 }\left [ 1-e^{-\eta ( 1+\eta ) \eta_{\rm d}\mu/2 } \right ],
\nonumber\\
\label{PnoDlinf:eq}
\eea
for large values of $L\gtrsim100$ km. This value is obtained from Eqs. (\ref{Pcon:eq}) and (\ref{Perr:eq}), by noting that the cosine term in Eq. (\ref{PnoDl:eq}) gives a zero contribution for sufficiently large values of  $\sigma_\phi$.
This finding shows the detrimental effect that the phase drift (and  the length of propagation) has on the protocol. Given that Alice and Bob discard the data associated with an inconclusive outcome (sifting), in Fig. \ref{Pcei:fig}(b) we plot the probability of correct outcome and the probability of 
error in the remaining $M_{\rm sift} \simeq M(1-p_{\rm inc})$ sifted data.  
The two probabilities approach 0.5 and for $L>150$ km the two curves have practically converged to this value.

When $P (\textrm{cor})>P (\textrm{err})$ Alice and Bob can establish a common secret binary key by applying  standard error correction  \cite{ECreview} and privacy amplification \cite{PA1,PA2} techniques, 
which are used extensively in QKD protocols and they are also applicable in the present framework. 
A detailed discussion goes beyond the aim of the present work, and the interested reader may refer to related papers in the literature \cite{ECreview,PA1,PA2}. 
However, in the following subsections the role of error correction and privacy amplification are taken into account.

 \begin{figure*}
\centerline{\includegraphics[width=1\linewidth]{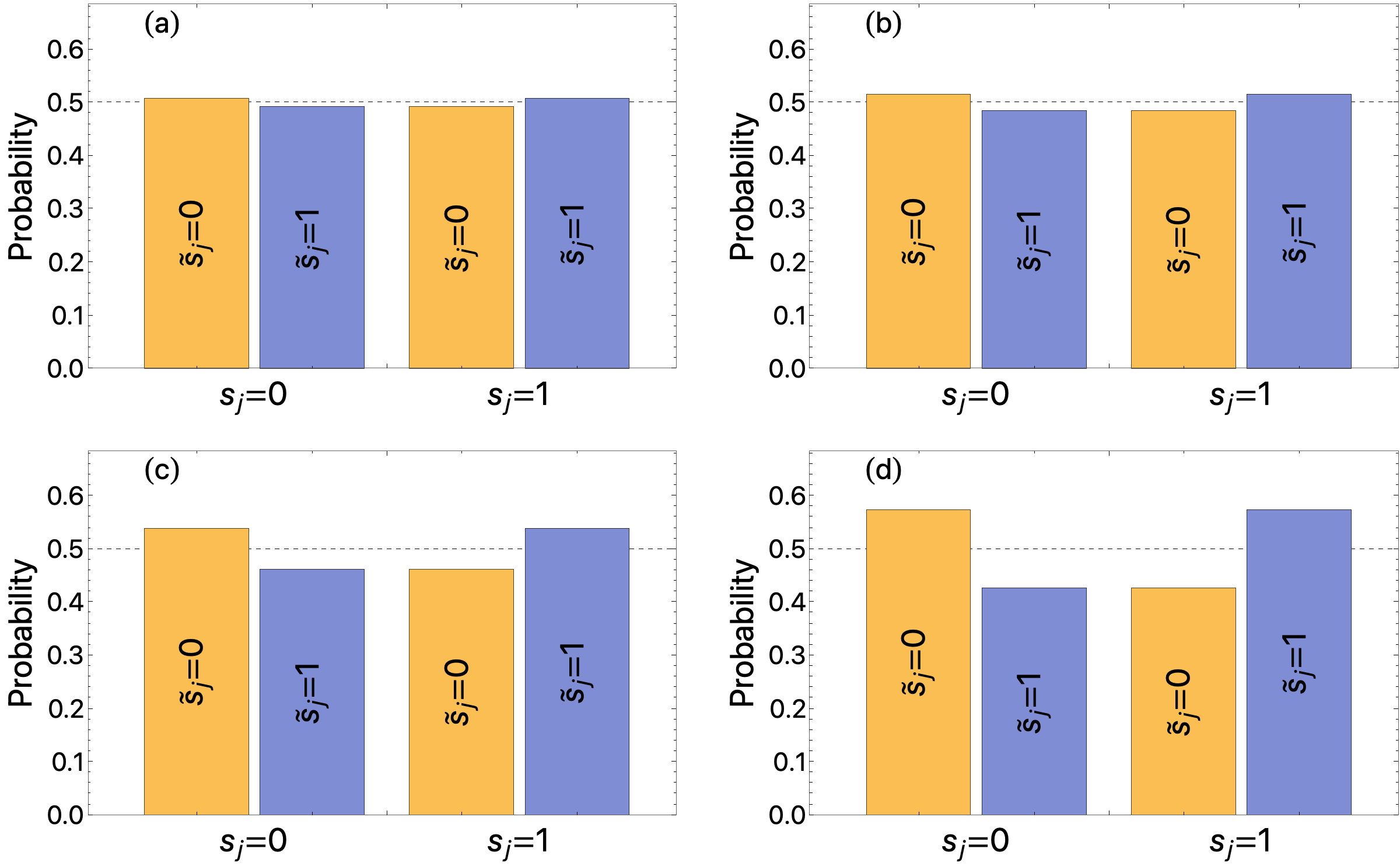}}
\caption{(Color online) (a) Conditional probability $P(\tilde{s}|s)$ for the eavesdropper to obtain $\tilde{s}$ given the actual bit value $s$, following the  attack discussed in the main text, and relies on minimum-error-discrimination measurements. 
Parameters: (a) $\mu = 0.01$, $N=20$. (b) $\mu = 0.02$, $N=20$. 
(c) $\mu = 0.05$, $N=20$. (b) $\mu = 0.1$, $N=20$. The adversary has a perfect channel $\eta = 1$, and perfect detectors $\eta_{\rm d} = 1$.} 
\label{PSs:fig}
\end{figure*}

\subsection{Minimum-error-discrimination attack} 
\label{sec3b}

Let us consider now  what Eve can do, in order to deduce the secret bit $s_j$. 
First of all, we assume that Eve can replace (without being detected), the imperfect channel that connects Alice and Bob, with a perfect one, in order to 
take advantage of the inevitable losses and imperfections. 
Eve has access to $\ket{\psi_{a_j}}$ and $\ket{\psi_{b_j}}$, when 
they are sent from Alice to Bob, and vice versa. 
She never has access to the quantum key $\ket{\psi_{k_j}}$, which is used for the encoding of the bit $s_j$. 
It is worth recalling here that, by virtue of the OTP encryption, the  key $k_j = a_j\oplus b_j$,  is information-theoretically secure provided that 
$a_j$ and $b_j$ are independent random uniformly distributed over $\Int_N$, and Eve does not have access to them.  
However, given that Eve has access to $\ket{\psi_{a_j}}$ and $\ket{\psi_{b_j}}$, she  can 
apply minimum-error measurements on the two states, 
in order deduce $a_j$ and $b_j$, respectively.  
Based on these estimates, say $\tilde{a}_j$ and $\tilde{b}_j$, 
she can make an educated guess about $k_j$, as follows 
$\tilde{k}_j = \tilde{a}_j\oplus \tilde{b}_j$. Subsequently, she prepares new states 
$\ket{\psi_{\tilde {a}_j}}$ and $\ket{\psi_{\tilde {b}_j}}$, which are compatible with her outcomes, and she sends them to Bob and Alice, 
respectively. In accordance with the discussion of Sec. \ref{sec2b}, throughout this section we 
focus on values for the parameters $\mu$ 
and $N$ where the map (\ref{map:eq})  behaves as a QOWF, 
in the sense that the probability for Eve 
to deduce correctly a transmitted state from the set (\ref{set_coh:eq})
is close to random guessing. 
Of course, Eve does not know whether her measurement has been successful or not, and thus she has to send to the users a state which is compatible with 
her outcomes. 
By contrast to QKD protocols,  Eve's intervention may 
introduce 
errors in the quantum keys of both Alice and Bob, because Eve has to measure both 
$\ket{\psi_{a_j}}$ and $\ket{\psi_{b_j}}$, in order to deduce $k_j$. 

The main question arises  is whether based on her estimates for $a_j$ and $b_j$ Eve can deduce the key bit $s_j$, or the bit is protected by the QOWF. 
Recall that Eve  prepares and sends to Alice and Bob the states 
$\ket{\psi_{\tilde{b}_j}}$ and  $\ket{\psi_{\tilde{a}_j}}$, respectively, which are in agreement with her estimates. 
Hence, due to Eve's actions, Alice encodes $s_j$ onto $\ket{\psi_{a_j\oplus \tilde{b}_j}}$ instead of $\ket{\psi_{a_j\oplus b_j}}$, as follows, 
\be
\hat{\cal U}^{s_j N/2}\ket{\psi_{a_j\oplus \tilde{b}_j}} := \ket{\psi_{c_j}},
\ee
which is in agreement with Eq. (\ref{encoding:eq}), for $c_j = a_j\oplus \tilde{b}_j \oplus s_jN/2$. 
This cipher state is sent to Bob, and can be intercepted 
by Eve, who applies a rotation $\hat{\cal U}^{-\tilde{a}_j\oplus \tilde{b}_j} $ thereby obtaining $\ket{\psi_{a_j\ominus \tilde{a}_j\oplus s_j N/2}}$.
Eve can deduce $s_j$ by measuring the X-quadrature of the resulting state: if the outcome  $x$ is positive she concludes $s_j=0$, otherwise $s_j =  1$. 
Note that Eve's outcome, depends only on $a_j\ominus \tilde{a}_j$, and not on $b_j$ or  $\tilde{b}_j $. Finally, Eve sends to Bob the state  
\bea
\ket{\psi_{\tilde{c}_j}} = \ket{\psi_{\tilde{a}_j\oplus \tilde{b}_j\oplus \tilde{s}_j N/2}},
\label{psi_cj:eq}
\eea
where $\tilde{s}_j$ is Eve's estimation. 
As depicted in Fig. \ref{PSs:fig}, the conditional probability $P(\tilde{s}_j|s_j)$ for Eve to obtain $\tilde{s}_j$ given that the input state is $s_j$, is symmetric in the sense that 
\bea
&&P(s_j|s_j) = P(s_j\oplus 1  | s_j\oplus 1),\nonumber\\
&&P(s_j\oplus 1|s_j) = P(s_j  | s_j\oplus 1),
\label{sym1:eq}
\eea
and thus it depends only on the difference $s_j\ominus \tilde{s}_j$. 
Most importantly, for sufficiently small values of $\mu$ (i.e., $\mu\lesssim 0.02$) the probabilities for Eve to obtain either 0 or 1, for a given $s_j\in\{0,1\}$, are approximately the same 
i.e., $P(\tilde{s}_j=0|s_j)\simeq P(\tilde{s}_j=1|s_j)\simeq 0.5$, which suggests that in this case, it is very hard for Eve to deduce the right value of $s_j$, because it is equally likely to obtain 0 or 1, for any given $s_j$. The difference  between $P(\tilde{s}_j=0|s_j)$ and $P(\tilde{s}_j=1|s_j)$ increases with increasing $\mu$, and Eve's ability to deduce the right bit improves. 
Hence, the following discussion will be mainly focused on $\mu\lesssim 0.02$, since in this case the encoded bit is almost perfectly protected by the QOWF.

Although Eve replaces the imperfect quantum channel with a perfect one, 
Eve's intervention is expected to introduce errors in the bit strings of Alice and Bob, because its strategy relies on the outcomes of two independent measurements, and at least one of them may be wrong. Alice and Bob have to sacrifice a number of bits in order to estimate the error rate in their strings, and based on this estimate, to decide on whether an attack has taken place. 
The main question is: What is the theoretically expected error rate in the presence of eavesdropping, especially for  $\mu\lesssim 0.02$?
As shown in appendix \ref{app2}, in the presence of eavesdropping  
the error probability in the  keys of Alice and Bob is given by 
\begin{widetext}
\bea
P({\rm err}) = \sum_{\tilde{s}_j} \sum_{\tilde{b}_j\ominus b_j} 
P(\textrm{one click at }D_{l=1}|\tilde{b}_j\ominus b_j,  \tilde{s}_j )p(\tilde{b}_j \ominus b_j)   p(\tilde{s}_j|s_j=0), 
\label{Perr:eq1}
\eea
where 
\bea
P(\textrm{one click at }D_l | \tilde{b}_j\ominus b_j, \tilde{s}_j ) &=& [1-P(\textrm{no click at }D_l | \tilde{b}_j, b_j, \tilde{s}_j ) ]
\times P(\textrm{no click at }D_{ l\oplus 1}| \tilde{b}_j, b_j, \tilde{s}_j ),
\label{P1ClickEve:eq}
\eea
and
\bea
&&P(\textrm{no click at }D_l | \tilde{b}_j, b_j, \tilde{s}_j ) 
=
\exp \left \{ 
-\eta_{\rm d}\mu
\left [ 1+(-1)^{l+\tilde{s}_j} \cos \left [ \frac{ 2(\tilde{b}_j \ominus b_j ) \pi }{N}\right  ]\right ] 
\right \},
\nonumber\\
\label{PnoDlattack:eq}
\eea 
which depends only on $\tilde{b}_j \ominus b_j $, and thus we write $P(\textrm{one click at }D_{l=1}|\tilde{b}_j\ominus b_j,  \tilde{s}_j )$ in Eq. (\ref{Perr:eq1}).
Working similarly for the probability of sharing the same bit value we obtain 
\bea
P(\textrm{cor})=
\sum_{\tilde{s}_j} \sum_{\tilde{b}_j\ominus b_j} 
P(\textrm{one click at }D_{l=0}|\tilde{b}_j\ominus b_j,  \tilde{s}_j )p(\tilde{b}_j \ominus b_j)   P(\tilde{s}_j|s_j=0).
\eea
\end{widetext}

\begin{figure}
\centerline{\includegraphics[width=1\linewidth]{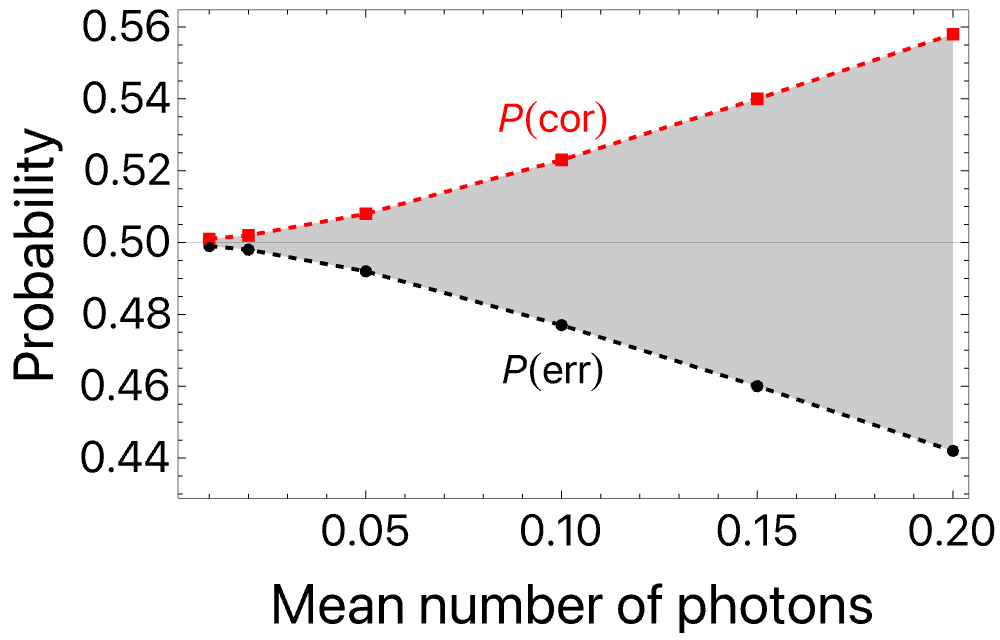}}
\caption{(Color online) Probabilities $P({\rm cor})$ and $P({\rm err})$ in the sifted  keys of Alice and Bob, as functions of the mean number of photons, in the presence of the eavesdropping attack discussed in the main text. 
The adversary replaces the imperfect channel with a perfect one, but she cannot control the detection efficiency 
at Alice's and Bob's sites Parameters: $N=20-30$, $\eta = 1$, $\eta_{\rm d} = 0.3$.} 
\label{PlotProbAttack:fig}
\end{figure}

 \begin{figure*}
\centerline{\includegraphics[width=1\linewidth]{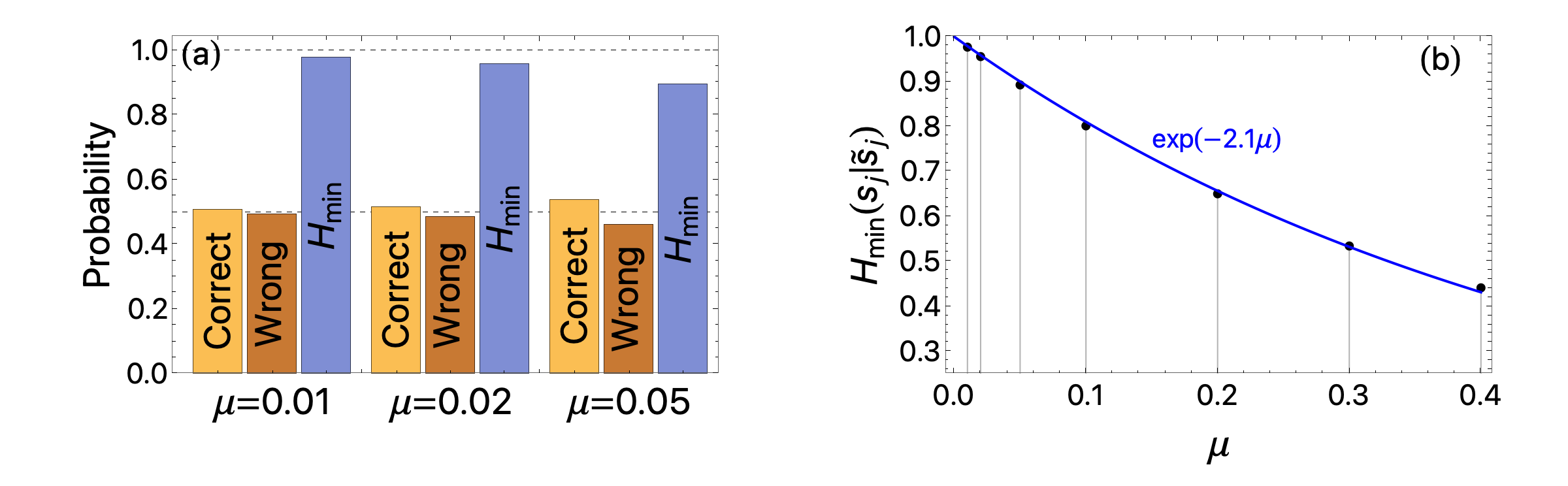}}
\caption{(Color online)  (a) The probability of correct and wrong guessing of the encoded bit value, and the corresponding min-entropy. Parameters: $\mu = 0.01$, $N=20$. The adversary has a perfect channel $\eta = 1$, and perfect detectors $\eta_{\rm d} = 1$. 
(b) The conditional min-entropy $H_{\rm min}(s_j|\tilde{s}_j)$, is plotted as a function of the mean number of photons $\mu$. The solid line is an exponential fit to the numerical data.} 
\label{Pmin:fig}
\end{figure*}

In view of the summation over $\tilde{s}_j$, the symmetry (\ref{sym2:eq}), and Eq. (\ref{P1ClickEve:eq}), one expects that 
$P(\textrm{err}) \simeq P(\textrm{cor})$ when $ P(\tilde{s}_j = 0|s_j=0)\simeq  P(\tilde{s}_j = 1|s_j=0) \simeq 0.5$, which occurs for $\mu\leq 0.02$ (see Fig. \ref{PSs:fig}). 
In this case, after ignoring the inconclusive outcomes,  one has $P(\textrm{err}) \simeq P(\textrm{cor})\simeq 0.5$ in the sifted key, which is confirmed by our simulations (see Fig. \ref{PlotProbAttack:fig}).

Summarizing the findings of the present and the previous sections, we have that for $\mu\lesssim 0.02$ and sufficiently large $N$, the quantum key $k_j$ and the encoded bit $s_j$ remain practically unknown to Eve, while her intervention results in very high error rates (around 0.5) in the sifted keys of Alice and Bob, which render error correction and privacy amplification processes highly inefficient, if at all possible. 
Recalling the findings and the discussion of Sec. \ref{sec3a}, in the absence of eavesdropping, so high error rates can be 
explained only in the framework of a very lossy channel with large random phase drift, corresponding to $L\gtrsim 100$ km. 
Of course, Alice and Bob cannot know if the observed  error rate is due to
innocent noise, or due to eavesdropping. Hence, they have to abort the protocol if the estimated error rate exceeds some value, say $P_{\rm max}^{\rm err}$. 
 In choosing $P_{\rm max}^{\rm err}$, one has to take into account 
the following facts. (i) For $\mu\lesssim 0.02$ (where the secrecy of the encoded bit is ensured), Eve's intervention yields error rates around 0.5. 
(ii) The error estimation by Alice and Bob will suffer from inevitable statistical deviations,  and one needs sufficiently large sample size,  in order to estimate the error rate with sufficient precision. 
(iii) Error correction becomes rather tedious and costly for high error rates, while error-rate thresholds below 0.2 or so, 
are found in standard QKD protocols, and thus related error-correction techniques are available. (iv) Small values of $P_{\rm max}^{\rm err}$ imply short propagation distances since, in the absence of eavesdropping, the error rate is associated with random phase drift, which increases with the distance (see related discussion in Sec. \ref{sec3a}). 
Based on all of these facts, we believe that  a choice of  $P_{\rm max}^{\rm err} = 0.2$ is reasonable for $\mu\lesssim 0.02$, because on the one hand an eavesdropping attack in the framework described here will yield considerably larger error rates $(\sim 0.5)$ and will be detected easily,  while on the other in the absence of 
eavesdropping attack, it corresponds to propagation distances of about 50 km (see Fig. \ref{Pcei:fig}), and it allows for the use of standard error-correction techniques.

Using the Chernoff bound, one can show \cite{NikDiaSciRep17} that a sample of size at least as large as 
\bea
M_{\rm th} := \frac{3\ln(2\xi^{-1})}{\varepsilon^{2}}, 
\label{Mth:eq}
\eea
is sufficient for Alice and Bob to be $100(1-\xi)\%$ confident that the estimated error rate $q$ lies within an interval of size $2\varepsilon$ 
around the actual error probability i.e., $|q - P({\rm err})|< \varepsilon$.   
The parameter $\xi\ll 1$ is the uncertainty, and $\varepsilon\ll P_{\rm max}^{\rm err}$ is the absolute 
error in the sampling, which quantifies the statistical deviations. The sample size scales logarithmically with $\xi$, which suggests that high confidence can be achieved without a significant increase of $M_{\rm th}$. Then, for 
$\varepsilon = 2\times 10^{-2}$ and $\xi=10^{-6}$ we have 
$M_{\rm th}\approx 10^5$. 
By measuring  $M_{\rm test}\geq M_{\rm th}$ test states, Alice and Bob obtain the empirical probability $q$, and they  proceed to error correction 
only if  $q+\varepsilon \leq P_{\rm max}^{\rm err}$, where $P_{\rm max}^{\rm err}=0.2$.  
This is because, in this case the actual error rate satisfies  $P({\rm err})< P_{\rm max}^{\rm err}$, which cannot be explained in the framework of the minimum-error discrimination attack discussed above. With high probability, such attack will result in much higher values for $q$,  around 0.5 for $\mu\lesssim 0.02$. 

In closing this section, it is worth emphasizing that the present analysis relies on the assumption that Eve replaces the quantum channel that connects Alice and Bob with an ideal channel $(\eta =1, \sigma_\phi = 0)$, 
she intercepts each transmitted state independently, and applies a minimum-error-discrimination measurement. 
Given that the states are prepared at random and independently by Alice and Bob, it is reasonable to 
assume that Eve does not gain anything by addressing collectively the transmitted states.  Moreover, there do exist other measurements that one may consider, which do not minimize 
the error probability for Eve, and thus they are expected to result in higher error rates. The same will happen if  Eve does not replace the imperfect quantum channel with a perfect one. 

\subsection{Photon-number-splitting attack} 
\label{sec3c}

Photon-number-splitting (PNS) attack is a simple, yet powerful attack against QKD protocols that rely on weak coherent pulses \cite{PNSAttack}. Eve exploits the inevitable losses in the channel in order to block one-photon pulses, and to extract a photon from each multiphoton pulse, thereby obtaining a copy of the actual state of the pulse. The attack does not introduce any errors, and can be detected only by means of decoy states. 

Interestingly enough, the PNS attack does not seem to be so useful against the  QDH protocol under consideration, and it is expected to introduce errors that will be detected by Alice and Bob. More precisely, 
Eve has to extract a photon from two independent weak coherent states, one that is sent from Alice to Bob $\ket{\psi_{a_j}}$, and one that is sent from Bob to Alice $\ket{\psi_{b_j}}$. Extracting a photon from one of these states only is not enough, because Eve's aim is to obtain a copy of the quantum-key state $\ket{\psi_{k_j}} = \ket{\psi_{a_j\oplus b_j}}$.  The probability for both $\ket{\psi_{a_j}}$ and $\ket{\psi_{b_j}}$ to contain at least two photons each, is at most $\mu^4/4$, which suggests that for $\mu\leq 0.1$, the probability for Eve to extract a photon from both states is at most $2.5\times 10^{-5}$. In these seldom cases, she will have copies of $\ket{\psi_{a_j}}$ and $\ket{\psi_{b_j}}$.  
Yet, this is not enough for Eve, because we are not aware of any technique for the generation of $\ket{\psi_{a_j\oplus b_j}}$ directly 
from $\ket{\psi_{a_j}}$ and $\ket{\psi_{b_j}}$ without measuring these states, thereby introducing errors in the keys of Alice and Bob. In other words, in the very rare cases that the PNS attack is possible, 
Eve has to measure $\ket{\psi_{a_j}}$ and $\ket{\psi_{b_j}}$ in order to obtain estimates of ${a}_j$  and ${b}_j$, and from these estimates, she can make an educated guess about $a_j\oplus b_j$, which can be used for the extraction of the  bit $s_j$ that is encoded on $\ket{\psi_{a_j\oplus b_j}}$. 
But this is precisely the attack discussed in the preceding section, which in fact 
yields estimates of ${a}_j$  and ${b}_j$ with the minimum possible error probability, because it relies on minimum-error-discrimination measurements. 
Hence, we conjecture that the best option for Eve to attack the proposed QDH protocol is the minimum-error-discrimination attack 
 considered in Sec. \ref{sec3b}. 

\subsection{Secret-key length}
\label{sec3d}

As depicted in Fig. \ref{Pmin:fig}(a), the average probability of correct guessing for Eve $p_{\rm E}({\rm cor})=\sum_{s_j}P(\tilde{s}_j=s_j|s_j)p(s_j)$,  increases with increasing $\mu$, while it does not depend on the number of phase 
slices for $N\geq N^\star$. The smaller $\mu$ is the closer the probability of correct guessing is to random guessing. 

The conditional min-entropy $H_{\min}({\bm s}|\tilde{\bm s})$, where $\tilde{\bm s}$ denotes Eve's guess about ${\bm s}$ based on measurement outcomes, is associated with the secure key length that can be extracted from ${\bm s}$ \cite{}. 
Taking into account that Alice and Bob prepare each state at random and independently, while Eve attacks each one of them independently following the same strategy, we have 
$H_{\min}({\bm s}|\tilde{\bm s}) = |{\bm s}| H_{\min}(s_j|\tilde{s}_j)$, where $|{\bm s}|$  is the length of the binary string shared between Alice and Bob, just before the error correction step.  
Moreover, taking into account the  information leak ${\mathscr L}$ during 
error correction and verification, one obtains that the 
number of secret bits $\ell$ that can be extracted from ${\bm s}$ and it is 
$\Delta-$close to uniform satisfies \cite{KeyRate1,KeyRate2,KeyRate3}
\bea
\ell \geq |{\bm s}| H_{\min}(s_j|\tilde{s}_j) - 
{\mathscr L}+2\log(\Delta),
\label{key_length:ineq}
\eea
where $\Delta \ll 1$. 
This expression shows clearly that the min-entropy plays pivotal role 
in the number of uniform independent random bits that can be extracted from 
${\bm s}$. 

The min-entropy $H_{\min}(s_j|\tilde{s}_j)$  is given by the negative logarithm of the optimal guessing probability \cite{KeyRate1,KeyRate2,KeyRate3}. 
In the eavesdropping strategy presented above, the minimum-error discrimination measurement that is used by Eve in order to deduce $a_j$ ensures that the probability for $ \tilde{a}_j=a_j$ 
is maximized, and for this reason we conjecture that the probability $p_{\rm E}({\rm cor})$ for Eve to deduce the correct value of $s_j$ is also maximal. 
Hence, we have  
\bea
H_{\min}(s_j|\tilde{s}_j) = -\log_2[p_{\rm E}({\rm cor})],
\eea
which is also shown in Fig. \ref{Pmin:fig}, for different values of the mean number of photons $\mu$. 
For $\mu\lesssim 0.02$, we have $H_{\min}\geq 0.96$, and it decays 
exponentially with increasing values of $\mu$  as $H_{\min} \approx e^{-2.1\mu}$, while it does not vary considerably with $N\geq N^\star$.
These findings suggest that for sufficiently small values of $\mu\lesssim 0.02$, 
and sufficiently large values of $N\geq N^\star$, the encoded bit $s_j$ is well protected by the QOWF, and it remains practically unknown to Eve for the attack under consideration.  
Hence, in inequality (\ref{key_length:ineq}) we have 
$|{\bm s}| H_{\min}(s_j|\tilde{s}_j)\simeq |{\bm s}|$, 
which is yet another manifestation of the role of the QOWF in the QDH scheme under consideration. 
Indeed, by choosing $\mu\geq 0.1$ and  irrespective of the number of slices, one will  increase the probability of conclusive outcomes, but at the same time 
the information that a potential eavesdropper can extract about ${\bm s}$ increases. Hence, the map (\ref{map:eq}) starts deviating from a QOWF and 
the encoded bit $s_j$ is not sufficiently protected any more. 
According to Fig. \ref{Pmin:fig}(b), $H_{\min}(s_j | \tilde{s}_j) \lesssim 0.8$ for  $\mu\geq 0.1$.

\section{Discussion and conclusions}
\label{sec4}

We have discussed an anonymous quantum DH (QDH) key-exchange 
protocol, which relies on the encoding of  integers in the phase of  
weak coherent states. 
We have investigated and identified the regime of parameters where the 
proposed mapping between integers and states may operate as a quantum 
one-way function. 
Moreover, a potential adversary has to attack pairs of independent random 
states, which are exchanged between Alice to Bob, thereby introducing errors 
at both Alice's and Bob's sites. 

In cryptography (classical or quantum), there are no protocols that solve all the cryptographic tasks simultaneously. 
Protocols for different tasks are discussed and analyzed separately, but two or more of them are combined judiciously, for the design of a secure and efficient cryptosystem.  
Similarly to the basic form of conventional DH key exchange, throughout  this work we focused on anonymous quantum key exchange, 
in the sense that there is no  authentication which ensures the identity of the users involved in the exchange, and it is  essential  for the prevention of a man-in-the-middle attack. 
In analogy to QKD protocols, one can use asymmetric cryptography, physical unclonable functions, or preshared keys to achieve authentication in the proposed QDH protocol  \cite{Abidin2013,Wang-etal21,NikFis24}.

The QDH we have discussed bares similarities with  twin-field QKD protocols \cite{LucYuaDynShi18,twin2,twin3,twin4}, as it relies on the interference of weak coherent states. 
For this reason, it is essential for the users to have reliable phase-control, and phase-stabilization techniques. 
The related techniques that are used in twin-field QKD protocols, can be also applied in the present context.  
However, there do exist major differences. First of all, the QDH  under consideration does not require the presence of a  third party, although it can be easily modified in this direction. 
Secondly, its security relies on a quantum one-way function, which implies that  the mean number of photons, and the number of phase slices have to be such that the probability for an adversary to deduce the correct state is close (within $\epsilon\ll 1$) to random guessing. There is no such requirement in QKD protocols. In fact,  the values of $\mu$ and $N$ used in most related implementations \cite{LucYuaDynShi18,twin2,twin3,twin4}, do not justify the presence of a quantum-one-way function. Thirdly, QKD protocols that rely on weak coherent pulses, are susceptible to a PNS attack, which can be detected by using decoy states.  As discussed in Sec. \ref{sec3c}, the PNS attack is not so useful 
in the QDH protocol under consideration,  while it is expected to introduce errors 
that will be detected by Alice and Bob.  
For these reasons we conjecture that the best strategy for Eve is to attack the  
protocol using the minimum-error-discrimination attack considered in Sec. \ref{sec3b}. 

Our simulations and results suggest that the proposed protocol does not have any special requirements in terms of channel losses, detection efficiency, mean number of photons, number of phase slices, etc. In particular, we note that there exist implementations of QKD protocols \cite{twin3} where $\mu = 0.02$ and $N=16$, and in these 
cases the presence of a QOWF is justified in all respects discussed in Sec. \ref{sec2}.
The main challenge in our protocol is that the quantum-key state 
$\ket{\psi_{k_j}}$ and the cipher state $\ket{\psi_{c_j}}$ have to impinge simultaneously on the beam splitter at  Bob's site in order for their comparison to take place. 
To this end, Alice and Bob  have to adjust accordingly the transmission and the arrival of the 
states $\ket{\psi_{a_j}}$ and $\ket{\psi_{c_j}}$ (e.g., using optical switches, fiber-loop delay lines if necessary).  
Hence, experimental realization of the proposed QDH protocol is within reach of current technology.


%

\begin{acknowledgments}
This research was co-funded by the European Union under the Digital Europe Program grant agreement number 101091504.
\end{acknowledgments}

\section*{AUTHOR DECLARATIONS}

\subsection*{Conflict of Interest}
The authors has no conflicts to disclose.

\subsection*{Data Availability Statement}

The data that support the findings of this study are available within the article.

\appendix

\section{Calculation of probabilities in the presence of eavesdropping}
\label{app2}

In view of Eq. (\ref{psi_cj:eq}), Eq. (\ref{omegaL:eq1}) reads in the presence of the attack discussed in Sec. \ref{sec3b}, 
\bea
\omega_{l} &=&\frac{\sqrt{\eta_{\rm d}} \psi_{\tilde{a}_j\oplus\tilde{b}_j\oplus\tilde{s}_jN/2}+(-1)^l \sqrt{\eta_{\rm d}} \psi_{\tilde{a}_j\oplus b_j}}{\sqrt{2}} 
\nonumber\\
& =&\sqrt{\frac{\eta_{\rm d}\mu}2} \left [ e^{{\rm i}(\tilde{a}_j\oplus\tilde{b}_j\oplus\tilde{s}_jN/2)\delta\varphi}+(-1)^l e^{{\rm i}(\tilde{a}_j\oplus b_j)\delta\varphi}\right ]
\nonumber\\
& =&2\sqrt{\frac{\eta_{\rm d}\mu}2} e^{{\rm i}(\xi_j+\zeta_j+l\pi+ 2\nu_j\pi)/2} \left | \cos \left (\frac{\xi_j-\zeta_j-l\pi}{2} \right )  \right |
\nonumber\\
\eea
where $\xi_j := (\tilde{a}_j\oplus \tilde{b}_j\oplus \tilde{s}N/2)\delta\varphi$ , $\zeta_j := (\tilde{a}_j\oplus b_j )\delta\varphi$, $\nu_j = 0$ for $\xi_j\oplus \zeta_j\leq N/2$ and 1 otherwise, while $l\in\{0,1\}$.  We have assumed that Eve, has replaced the channel that connects Alice and Bob with a perfect channel, in order to take advantage of the expected losses. Hence, we have set $\eta = 1$. 

The probability for obtaining no click at detector $D_l$ is given by 
\bea
&&P(\textrm{no click at }D_l | \tilde{b}_j, b_j, \tilde{s}_j ) = |\langle 0 | \omega_l\rangle|^2 
\nonumber\\
&&=\exp \left \{ 
-\eta_d\mu
\left [ 1+(-1)^{l+\tilde{s}_j} \cos \left [ \frac{ 2(\tilde{b}_j \ominus b_j ) \pi }{N}\right  ]\right ] 
\right \},
\eea 
which is Eq. (\ref{PnoDlattack:eq}). It is worth noting here that this probability satisfies the relation 
\bea
&&P(\textrm{no click at }D_l | \tilde{b}_j, b_j, \tilde{s}_j) =P(\textrm{no click at }D_{l\oplus 1} | \tilde{b}_j, b_j,  \tilde{s}_j\oplus 1) 
\label{sym2:eq}
\nonumber\\
\eea

The probability for Bob to obtain a click only at the detector $D_l$ (in which case we have a conclusive outcome), is given by  
\begin{widetext}
\bea
P(\textrm{one click at }D_l | \tilde{b}_j, b_j, \tilde{s}_j ) &=& [1-P(\textrm{no click at }D_l | \tilde{b}_j, b_j, \tilde{s}_j ) ]
\times P(\textrm{no click at }D_{ l\oplus 1}| \tilde{b}_j, b_j, \tilde{s}_j ),
\eea
\end{widetext}
and Bob concludes that $s_j = l$.


The probability of error is given by
\begin{widetext}
\bea
P(\textrm{err})&=&
\sum_{s_j,\tilde{s}_j} \sum_{b_j,\tilde{b}_j}  \sum_{a_j,\tilde{a}_j} 
P(\textrm{one click at }D_{l\neq s_j},s_j , \tilde{s}_j, b_j, \tilde{b}_j, a_j, \tilde{a}_j)
\nonumber\\
&=& 
\sum_{s_j,\tilde{s}_j} \sum_{b_j,\tilde{b}_j}  \sum_{a_j,\tilde{a}_j} 
P(\textrm{one click at }D_{l\neq s_j}|b_j, \tilde{b}_j, \tilde{s}_j )p(s_j,\tilde{s}_j, b_j, \tilde{b}_j, a_j,\tilde{a}_j)
\nonumber\\
&=& 
\sum_{s_j,\tilde{s}_j} \sum_{b_j,\tilde{b}_j}  \sum_{a_j,\tilde{a}_j} 
P(\textrm{one click at }D_{l\neq s_j}|b_j, \tilde{b}_j,  \tilde{s}_j )p(b_j,\tilde{b}_j) p(s_j,\tilde{s}_j,a_j,\tilde{a}_j)
\nonumber\\
&=& 
\sum_{s_j,\tilde{s}_j} \sum_{b_j,\tilde{b}_j} 
P(\textrm{one click at }D_{l\neq s_j}|b_j, \tilde{b}_j,  \tilde{s}_j )p(b_j,\tilde{b}_j)   p(s_j,\tilde{s}_j)
\nonumber\\
&=& 
\frac{1}{2N}\sum_{s_j,\tilde{s}_j} \sum_{b_j,\tilde{b}_j} 
P(\textrm{one click at }D_{l\neq s_j}|b_j, \tilde{b}_j,  \tilde{s}_j )P(\tilde{b}_j|b_j)   P(\tilde{s}_j|s_j)
\nonumber\\
&=& 
\frac{1}{2N}\sum_{s_j,\tilde{s}_j} \sum_{b_j} \sum_{\tilde{b}_j\ominus b_j} 
P(\textrm{one click at }D_{l\neq s_j}|\tilde{b}_j\ominus b_j,  \tilde{s}_j )P(\tilde{b}_j \ominus b_j|b_j)   P(\tilde{s}_j|s_j)
\nonumber\\
&=& 
\frac{1}{2}\sum_{s_j,\tilde{s}_j} \sum_{\tilde{b}_j\ominus b_j} 
P(\textrm{one click at }D_{l = s_j\oplus 1}|\tilde{b}_j\ominus b_j,  \tilde{s}_j )p(\tilde{b}_j \ominus b_j)   P(\tilde{s}_j|s_j)
\nonumber\\
&=& 
\sum_{\tilde{s}_j} \sum_{\tilde{b}_j\ominus b_j} 
P(\textrm{one click at }D_{l= 1}|\tilde{b}_j\ominus b_j,  \tilde{s}_j )p(\tilde{b}_j \ominus b_j)   P(\tilde{s}_j|s_j=0).
\eea
\end{widetext}

In the second equality we have used the fact that the probability of click in a detector is independent of $a_j$ and $\tilde{a}_j$, while  the third equality holds because 
$b_j$ and $\tilde{b}_j$ are independent of $a_j$, $\tilde{a}_j$, $s_j$ and $\tilde{s}_j$. In the fourth equality we have performed the summation over $a_j$ and $\tilde{a}_j$, 
using the fact that they are independent of  $s_j$ and $\tilde{s}_j$. In the fifth equality the joint probabilities have been expressed in terms of conditional probabilities, while 
 $p(b_j)=1/N$ and $p(s_j)=1/2$. According to Eq. (\ref{PnoDlattack:eq}), the probability depends on $p(\tilde{b}_j\ominus b_j)$ while 
 our simulations suggest that $P(\tilde{b}_j\ominus b_j|b_j)$ is independent of $b_j$. These facts bring us to the sixth and the seventh equality.  Finally, 
 using Eqs. (\ref{sym2:eq}) and  (\ref{sym1:eq}), we have performed 
 the summation over $s_j$  in the last equation.

\end{document}